\documentclass{article}
\usepackage{authblk}
\usepackage[sorting=none]{biblatex}
\usepackage{enumitem}
\usepackage{hyperref}
\usepackage{amsmath}
\usepackage{amsfonts}
\usepackage{pgfplots}
\usepackage{pgfplotstable}
\usepackage{float}
\pgfplotsset{compat=1.17}
\usepgfplotslibrary{dateplot}
\usepackage{eurosym}
\usepackage{subcaption}
\usepackage[braket, qm]{qcircuit}
\usepackage{graphicx}
\title{Exploring Hybrid Quantum-Classical Methods for Practical Time-Series Forecasting}
\author[1,*]{Maksims~Dimitrijevs}
\author[1,*]{M\={a}rti\c{n}\v{s}~K\={a}lis}
\author[1,*]{I\c{l}ja~Repko}
\affil[1]{Centre for Quantum Computer Science, Faculty of Sciences and Technology, University of Latvia}
\affil[*]{Project conducted in collaboration with Forse AI (Sapiens.BI SIA)}

\begin{document}

\maketitle

\begin{abstract}
  Time-series forecasting is essential for strategic planning and resource allocation. In this work, we explore two quantum-based approaches for time-series forecasting. The first approach utilizes a Parameterized Quantum Circuit (PQC) model. The second approach employs Variational Quantum Linear Regression (VQLS), enabling time-series forecasting by encoding the problem as a system of linear equations, which is then solved using quantum optimization techniques. We compare the results of these two methods to evaluate their effectiveness and potential advantages for practical forecasting applications.
\end{abstract}

\section{Introduction}

Time-series forecasting is a critical tool in fields, where accurate predictions of future values based on past observations can drive strategic decision-making and resource optimization. The fundamental objective in time-series forecasting is to develop a model that effectively captures underlying patterns in historical data and extrapolates them to future points. One classical approach to this problem involves framing the prediction as a regression task, where a sliding window of previous time-steps is used to forecast the next value in the series. This approach naturally leads to a linear system, where past values in the time series are combined through a set of coefficients to produce the forecast. Solving this system by determining the optimal coefficients allows for efficient predictions, and thus, the time-series forecasting task becomes closely aligned with finding solutions to linear equations --- a process that is computationally intensive for large datasets but offers promising advantages when adapted to quantum computing techniques.

Quantum computing harnesses the principles of quantum mechanics to process information. Unlike classical computers that use bits (0 or 1), quantum computers utilize qubits. Qubits can exist in a superposition, representing both 0 and 1 simultaneously. This fundamental difference allows quantum computers to perform computations in parallel, potentially leading to significant speedups for certain problem classes. However, one has to measure the quantum state to learn any results, and the \emph{interesting} results are only measured with some probability. The goal of quantum algorithms is to increase the probability that the \emph{interesting} results are measured with high enough probability.

While the potential of quantum computing is very high, current technology faces limitations. Currently we are in the era of Noisy Intermediate-Scale Quantum (NISQ) devices \cite{preskill2018quantum}. These devices are susceptible to noise and have limited qubit counts, hindering large-scale calculations. However, even with these constraints, researchers are exploring algorithms and applications suitable for NISQ devices, such as hybrid quantum-classical approaches.

Algorithms such as the Harrow–Hassidim–Lloyd (HHL) algorithm offer exponential speed-up over the best known classical algorithms, but require large and fault-tolerant quantum computer not available currently \cite{harrow2009quantum}. Hybrid quantum-classical algorithms like the Variational Quantum Linear Solver (VQLS) \cite{bravo-prieto_variational_2023} aim to mitigate the limitations of NISQ devices. These algorithms combine short-depth quantum circuits for cost function evaluation with classical optimization techniques, reducing the reliance on extensive quantum resources.

The performance of both HHL and VQLS algorithms depend on the condition number of the coefficient matrix of the linear system. There are no guarantees for the coefficient matrices of time-series problems to be well-conditioned. In fact, they are likely to have high condition numbers leading to weak performance guarantees. However, many models in machine learning do not have strong theoretical performance guarantees, but perform well in practice nonetheless. In this paper we set out to test the practical performance of a quantum version of a linear system solver.

\section{Methods}

The study evaluated the performance of a Parameterized Quantum Circuit (PQC) model \cite{benedetti2019parameterized}, where two optimization algorithms, L-BFGS-B and COBYLA, were employed to find optimal parameters. The PQC's performance was compared against two classical baselines: a linear regression model and a simple deep learning model. The deep learning model consisted of two hidden layers, each with 12 units and ReLU activation, to mirror the limitations of the currently available quantum devices. The primary evaluation metric used across all models was the Mean Squared Error (MSE).

The dataset used in the experiments was provided by \emph{Forse AI}. It is a synthetic dataset designed to mimic typical sales data patterns analyzed by the company.

Standard data preprocessing steps were applied to ensure the data was suitable for both classical and quantum models. To address non-stationarity often present in time-series sales data, differencing was performed, which enhanced the stationarity of the dataset. This step is critical for improving the performance of models that assume stationarity in their inputs. Furthermore, the data was scaled to fall within a normalized range, a necessary condition to facilitate encoding on a quantum device, which operates effectively only on appropriately scaled data. This preprocessing ensures compatibility and optimal performance across all evaluated models.

\begin{figure}[hbt!]
    \centering
    
    \begin{subfigure}{0.9\textwidth}
        \centering
        \begin{tikzpicture}
            \definecolor{verylightgray}{rgb}{0.9, 0.9, 0.9}
            \begin{axis}[
                width=\textwidth,
                height=0.4\textwidth,
                date coordinates in=x,
                xticklabel style={rotate=45, anchor=east, font=\small},
                xlabel={Date},
                xlabel style={at={(1,-0.1)}, anchor=north east},
                ylabel={Sales (M\euro)},
                grid=major,
                axis lines=left,
                hide x axis=false,
                hide y axis=false,
                clip mode=individual,
                xtick={2019-01-01,2020-01-01,2021-01-01,2022-01-01,2023-01-01},
                xticklabel={\year-\month},
                tick align=outside,
                scaled y ticks=false,
                ymin=0,
                axis on top
            ]
                \addplot[draw=none, fill=verylightgray] 
                    coordinates {(2019-01-01,0) (2021-09-01,0) (2021-09-01,6) (2019-01-01,6)};
                \addplot[
                    mark=*,
                    mark size=1pt
                ]
                table[
                    x=Date,
                    y expr=\thisrow{Sales}/1000000,
                    col sep=comma
                ] {data/sales.csv};
            \end{axis}
        \end{tikzpicture}
        \caption{Original data. Data in the shaded area is used for training and scaling adjustments The rest is used for testing.}
        \label{fig:original-sales}
    \end{subfigure}

    \vspace{0.5cm} 

    \begin{subfigure}{0.9\textwidth}
        \centering
        \begin{tikzpicture}
            \definecolor{verylightgray}{rgb}{0.9, 0.9, 0.9}
            \begin{axis}[
                width=\textwidth,
                height=0.4\textwidth,
                date coordinates in=x,
                xticklabel style={rotate=45, anchor=east, font=\small},
                xlabel={Date},
                xlabel style={at={(1,-0.1)}, anchor=north east},
                ylabel={Preprocessed sales},
                grid=major,
                axis lines=left,
                hide x axis=false,
                hide y axis=false,
                clip mode=individual,
                xtick={2019-01-01,2020-01-01,2021-01-01,2022-01-01,2023-01-01},
                xticklabel={\year-\month},
                tick align=outside,
                scaled y ticks=false,
                ymin=-0.5,
                ymax=0.5,
                axis on top
            ]
                \addplot[draw=none, fill=verylightgray] 
                    coordinates {(2019-01-01,-0.5) (2021-09-01,-0.5) (2021-09-01,0.5) (2019-01-01,0.5)};
                \addplot[
                    mark=*,
                    mark size=1pt
                ]
                table[
                    x=Date,
                    col sep=comma
                ] {data/sales_preprocessed.csv};
            \end{axis}
        \end{tikzpicture}
        \caption{Preprocessed data after applying differencing (change in sales vs. absolute sales) and scaling the differences to the range [-0.25, +0.25]. Note that test-time data can fall outside of this range.}
        \label{fig:preprocessed-sales}
    \end{subfigure}

    \caption{Comparison of monthly sales (M\euro) before and after preprocessing from 2019 to 2023.}
    \label{fig:comparison-sales}
\end{figure}
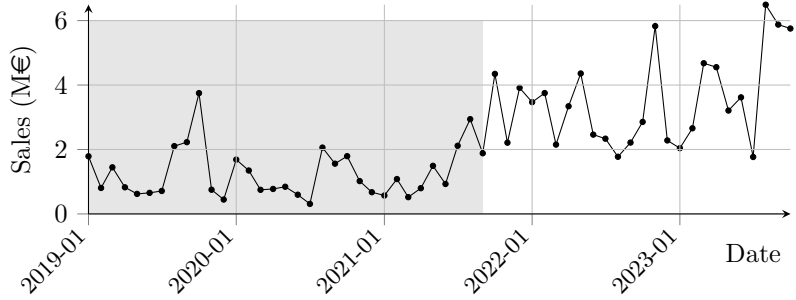
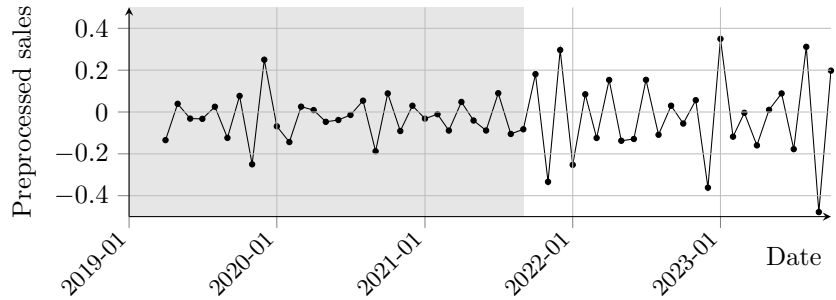
\pagebreak
\subsection{Parameterized Quantum Circuit}

The core of the algorithm lies in the PQC, which consists of two primary components:
\begin{itemize}
    \item \emph{Feature map} encodes the features of each training sample as angles of rotation gates applied to the qubits. In the experiments 12 previous time-steps are used as features, encompassing information on a full year.
    \item \emph{Parameterized ansatz} is comprised of entangling gates (CNOT in this case) and parameterized single-qubit gates ($R_X$ and $R_Y$). The parameters within these gates are what the algorithm adjusts during training to align the predicted values with the actual values from the training set.
\end{itemize}

The algorithm learns by iteratively updating the parameters of the ansatz to minimize the squared difference between the expected value of an observable (calculated using a Qiskit simulator) and the actual labels for each training samples.

\begin{figure}
    \centering
    \scalebox{0.8}{
    \Qcircuit @C=1.0em @R=0.2em @!R { \\
    	 	\nghost{\ket{{q}_{0}}} & \lstick{\ket{0}} & \gate{\mathrm{R_Y}(\phi_0)} & \ctrl{1} & \gate{\mathrm{R_X(\theta_{0,0})}} & \gate{\mathrm{R_Y(\theta_{0,1})}} & \qw & \targ & \gate{\mathrm{R_X(\theta_{1,0})}} & \gate{\mathrm{R_Y(\theta_{1,1})}} & \qw &\meter\\
    	 	\nghost{\ket{{q}_{1}}} & \lstick{\ket{0}} & \gate{\mathrm{R_Y}(\phi_1)} & \targ & \gate{\mathrm{R_X(\theta_{0,2})}} & \gate{\mathrm{R_Y(\theta_{0,3})}} & \ctrl{1} & \qw & \gate{\mathrm{R_X(\theta_{1,2})}} & \gate{\mathrm{R_Y(\theta_{1,3})}} & \qw &\meter\\
    	 	\nghost{\ket{{q}_{2}}} & \lstick{\ket{0}} & \gate{\mathrm{R_Y}(\phi_2)} & \ctrl{1} & \gate{\mathrm{R_X(\theta_{0,4})}} & \gate{\mathrm{R_Y(\theta_{0,5})}} & \targ & \qw & \gate{\mathrm{R_X(\theta_{1,4})}} & \gate{\mathrm{R_Y(\theta_{1,5})}} & \qw &\meter\\
    	 	\nghost{\ket{{q}_{3}}} & \lstick{\ket{0}} & \gate{\mathrm{R_Y}(\phi_3)} & \targ & \gate{\mathrm{R_X(\theta_{0,6})}} & \gate{\mathrm{R_Y(\theta_{0,7})}} & \ctrl{1} & \qw & \gate{\mathrm{R_X(\theta_{1,6})}} & \gate{\mathrm{R_Y(\theta_{1,7})}} & \qw &\meter\\
    	 	\nghost{\ket{{q}_{4}}} & \lstick{\ket{0}} & \gate{\mathrm{R_Y}(\phi_4)} & \ctrl{1} & \gate{\mathrm{R_X(\theta_{0,8})}} & \gate{\mathrm{R_Y(\theta_{0,9})}} & \targ & \qw & \gate{\mathrm{R_X(\theta_{1,8})}} & \gate{\mathrm{R_Y(\theta_{1,9})}} & \qw &\meter\\
    	 	\nghost{\ket{{q}_{5}}} & \lstick{\ket{0}} & \gate{\mathrm{R_Y}(\phi_5)} & \targ & \gate{\mathrm{R_X(\theta_{0,10})}} & \gate{\mathrm{R_Y(\theta_{0,11})}} & \ctrl{1} & \qw & \gate{\mathrm{R_X(\theta_{1,10})}} & \gate{\mathrm{R_Y(\theta_{1,11})}} & \qw &\meter\\
    	 	\nghost{\ket{{q}_{6}}} & \lstick{\ket{0}} & \gate{\mathrm{R_Y}(\phi_6)} & \ctrl{1} & \gate{\mathrm{R_X(\theta_{0,12})}} & \gate{\mathrm{R_Y(\theta_{0,13})}} & \targ & \qw & \gate{\mathrm{R_X(\theta_{1,12})}} & \gate{\mathrm{R_Y(\theta_{1,13})}} & \qw &\meter\\
    	 	\nghost{\ket{{q}_{7}}} & \lstick{\ket{0}} & \gate{\mathrm{R_Y}(\phi_7)} & \targ & \gate{\mathrm{R_X(\theta_{0,14})}} & \gate{\mathrm{R_Y(\theta_{0,15})}} & \ctrl{1} & \qw & \gate{\mathrm{R_X(\theta_{1,14})}} & \gate{\mathrm{R_Y(\theta_{1,15})}} & \qw &\meter\\
    	 	\nghost{\ket{{q}_{8}}} & \lstick{\ket{0}} & \gate{\mathrm{R_Y}(\phi_8)} & \ctrl{1} & \gate{\mathrm{R_X(\theta_{0,16})}} & \gate{\mathrm{R_Y(\theta_{0,17})}} & \targ & \qw & \gate{\mathrm{R_X(\theta_{1,16})}} & \gate{\mathrm{R_Y(\theta_{1,17})}} & \qw &\meter\\
    	 	\nghost{\ket{{q}_{9}}} & \lstick{\ket{0}} & \gate{\mathrm{R_Y}(\phi_9)} & \targ & \gate{\mathrm{R_X(\theta_{0,18})}} & \gate{\mathrm{R_Y(\theta_{0,19})}} & \ctrl{1} & \qw & \gate{\mathrm{R_X(\theta_{1,18})}} & \gate{\mathrm{R_Y(\theta_{1,19})}} & \qw &\meter\\
    	 	\nghost{\ket{{q}_{10}}} & \lstick{\ket{0}} & \gate{\mathrm{R_Y}(\phi_{10})} & \ctrl{1} & \gate{\mathrm{R_X(\theta_{0,20})}} & \gate{\mathrm{R_Y(\theta_{0,21})}} & \targ & \qw & \gate{\mathrm{R_X(\theta_{1,20})}} & \gate{\mathrm{R_Y(\theta_{1,21})}} & \qw &\meter\\
    	 	\nghost{\ket{{q}_{11}}} & \lstick{\ket{0}} & \gate{\mathrm{R_Y}(\phi_{11})} & \targ & \gate{\mathrm{R_X(\theta_{0,22})}} & \gate{\mathrm{R_Y(\theta_{0,23})}} & \qw & \ctrl{-11} & \gate{\mathrm{R_X(\theta_{1,22})}} & \gate{\mathrm{R_Y(\theta_{1,23})}} & \qw &\meter\\
    \\ }}
    \caption{Quantum Circuit with parameterized rotation gates. The first layer of gates encode the data by performing a rotation $\mathrm{R_Y}(\phi_i)$ by the preprocessed data in Figure~\ref{fig:preprocessed-sales}. $\phi_i$ are non-trainable. The next two layers entangle the data using two-qubit CNOT operations and learns parameter $\theta_{i,j}$ values that minimize the loss function.}
    \label{fig:quantum-circuit}
\end{figure}
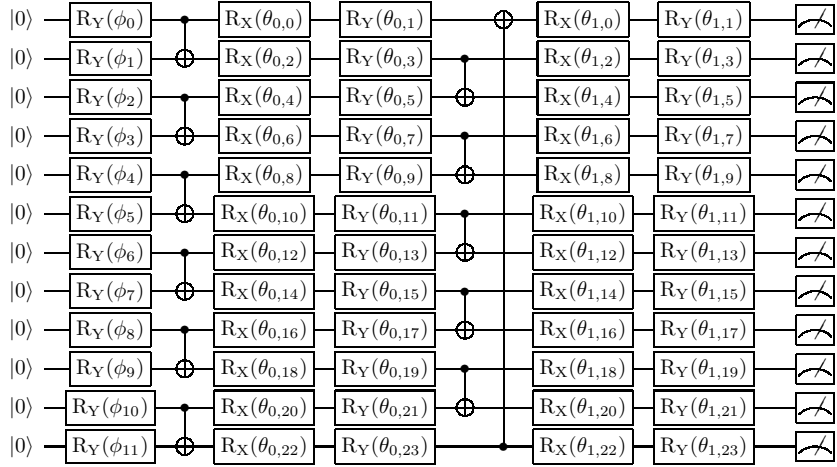

\subsection{Classical regression for time-series prediction}
\label{sec:time-series-reg}

To set up a time-series prediction problem using linear regression, suppose we have a time series with $n$ data points, represented as:

$$
\{y_t\}_{t=0}^{n-1} = \{y_0, y_1, y_2, \dots, y_{n-1}\}
$$

We want to use a sliding window approach to construct a matrix $X$ and a vector $y$ for supervised learning, where each row in $A$ represents a sequence of past time steps (or ``features'') used to predict the next time step in $y$ (the ``label'').

\subsubsection*{Step 1: Constructing the matrix $X$ and vector $y$}

\begin{enumerate}
    \item \textbf{Choose a window size} $m$, which determines how many time steps we use to predict the next step.

    \item \textbf{Form matrix $A$}: Each row in $A$ is a sequence of $m$ consecutive values from the time series, and there will be $n - m$ rows in total. Formally, $X \in \mathbb{R}^{(n - m) \times m}$ is given by:

    $$
    X = 
    \begin{bmatrix}
    y_0 & y_1 & \cdots & y_{m-1} \\
    y_1 & y_2 & \cdots & y_{m} \\
    y_2 & y_3 & \cdots & y_{m+1} \\
    \vdots & \vdots & \ddots & \vdots \\
    y_{n-m-1} & y_{n-m} & \cdots & y_{n-2}
    \end{bmatrix}
    $$

    \item \textbf{Form vector $y$}: For each row in $X$, the corresponding entry in $y$ is the next time step in the sequence. Thus, $y \in \mathbb{R}^{(n - m)}$ is given by:

    $$
    y = 
    \begin{bmatrix}
    y_{m} \\
    y_{m+1} \\
    y_{m+2} \\
    \vdots \\
    y_{n-1}
    \end{bmatrix}
    $$
\end{enumerate}

\subsubsection*{Step 2: Setting Up the Linear System $Xw = y$}

The goal is to find a vector $w \in \mathbb{R}^m$ that represents the coefficients used to linearly combine the values in each row of $X$ to predict the corresponding value in $y$. This setup leads to the linear system:

$$
Xw = y
$$

where:
\begin{itemize}[noitemsep]
    \item $X$ is the matrix of input sequences (past values).
    \item $w$ is the vector of coefficients (weights) that we want to determine.
    \item $y$ is the vector of target values (next time steps in the time series).
\end{itemize}

Each element $w_i$ in the vector $w$ represents the weight or influence of the $i$-th time step in the input window on the prediction of the next time step. Once $w$ is determined, it can be used to make predictions by applying it to new sliding windows from the time series.

Since $X$ is not guaranteed to be invertible, we can set $A = X^T X$ and $b = X^T y$, and use the normal equation $Aw = b$ to solve for $w$. Vector $w$ that is a solution to this system will minimize the squared error between $Xw$ and $y$, providing the best linear combination of previous time steps for predicting the next time step.

\subsection{Variational Quantum Linear Regression}

In quantum computing, vectors need to be represented as quantum states, which must be normalized. Thus, we need to:
\begin{itemize}
    \item Normalize the vector $b$ so that it can be represented as a quantum state $|b\rangle$.
    \item Reformulate the task as finding a $|w\rangle$ (the quantum state corresponding to the vector $w$) such that $A |w\rangle$ is proportional to $|b\rangle$. This implies that $A |w\rangle = \lambda |b\rangle$ for some scalar $\lambda$.
\end{itemize}

\subsubsection{Construction of matrix \texorpdfstring{$M$}{M} and vector \texorpdfstring{$b$}{b}}

Any matrix \( M \) acting on \( n \) qubits can be expressed as a linear combination of matrices \( M_i \) with coefficients \( \alpha_i \) \cite{OppenheimReznik} as follows:
\[
M = \sum_{i=0}^{4^n - 1} \alpha_i M_i,
\]
where $\alpha_i=\frac{\mathrm{Tr}(M_i \cdot M)}{2^n}$, and \( M_i \) is defined as the tensor product of Pauli matrices, \( \mathcal{P}_{ij} \), for each qubit:
\[
M_i = \bigotimes_{j=0}^{n-1} \mathcal{P}_{ij}.
\]
The Pauli matrices are given by:
\[
\begin{array}{cccc}
    \sigma_0 = 
    \begin{pmatrix}
    1 & 0 \\
    0 & 1
    \end{pmatrix}, & 
    \sigma_1 = 
    \begin{pmatrix}
    0 & 1 \\
    1 & 0
    \end{pmatrix}, & 
    \sigma_2 = 
    \begin{pmatrix}
    0 & -i \\
    i & 0
    \end{pmatrix}, & 
    \sigma_3 = 
    \begin{pmatrix}
    1 & 0 \\
    0 & -1
    \end{pmatrix}.
\end{array}
\]
The key idea is that \( M \) can be represented as a linear combination of the matrices \( M_i \), with the basis \( M_i \) being all possible combinations of tensor products of the Pauli matrices. Formally, each \( \mathcal{P}_{ij} \) is determined by the \( j \)-th digit of \( i \) in its quaternary (base-4) representation. Specifically, if \( i \) is written in base-4 as \( i = c_{n-1} c_{n-2} \cdots c_1 c_0 \), where \( c_j \in \{0, 1, 2, 3\} \), then $\mathcal{P}_{ij} = \sigma_{c_j}$.

However, in reality, we need at most $n^2$ non-zero terms of $M$ \cite{Werner}.

For instance, for \( i = 6 \) and \( n = 7 \), the base-4 representation of \( i \) is \( 0000012 \). Therefore:
\[
\mathcal{P}_{60} = \mathcal{P}_{61} = \mathcal{P}_{62} = \mathcal{P}_{63} = \mathcal{P}_{64} = \sigma_0, \quad 
\mathcal{P}_{65} = \sigma_1, \quad 
\mathcal{P}_{66} = \sigma_2.
\]

Vector $b$ is normalized to be represented as a quantum state $\ket{b}$. To obtain $\ket{b}$ from $b$, each element of $b$ should be divided by summation of squares of elements of $b$. We initialize quantum state $\ket{b}$ by using Qiskit initialization functionality, which resembles transpilation of quantum gates.

\subsubsection{\texorpdfstring{Reformulating the Task as Finding a State Vector $\ket{w}$}{Reformulating the Task as Finding a State Vector w}}

As discussed in Section~\ref{sec:time-series-reg}, for input $X$ and $y$, we aim to solve the equation for the vector $w$:
\[
    A w = b
\]
where $A = X^T X$ and $b = X^T y$. In the quantum setting, the solution for $w$ is typically found in terms of a unit vector:
\[
    M \ket{w} = \ket{b}
\]
where $\ket{w} = \frac{w}{\|w\|}$ and $\ket{b} = \frac{b}{\|b\|}$, and $M$ represents the matrix decomposition of $A$ in terms of Pauli operators, i.e., $M = \sum_i \alpha_i M_i$.

The problem is that we cannot directly compute $w$ from the quantum state $\ket{w}$, as the algorithm only provides the unit vector $\ket{w}$, and the norm $\|w\|$ is unknown. Although the vector $\ket{w}$ will be solved by VQLS using the decomposition of $A$, to find $w$, we can rescale the answer $\ket{w}$ using the following relation
\[
    \frac{A \ket{w}}{\|A \ket{w}\|} \approx \ket{b}
\]
Multiplying both sides of this equation by $\|b\|$, we get $b$ on the right side:
\[
    A \cdot \underbrace{\frac{\|b\|}{\|A \ket{w}\|} \ket{w}}_{w} \approx b
\]
Thus, the solution for $w$ is:
\[
    w \approx \frac{\|b\|}{\|A \ket{w}\|} \ket{w}
\]
,where $\ket{w}$ is the result obtained from the VQLS and $A$ input matrix.

\subsubsection{Parametrized Ansatz}

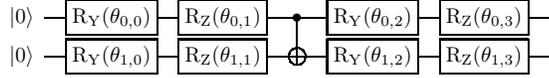
\begin{figure}[hbt!]
    \centering
    \scalebox{0.8}{
    \Qcircuit @C=1.0em @R=0.2em @!R { \\
    	 	\nghost{\ket{{q}_{0}}} & \lstick{\ket{0}} & \gate{\mathrm{R_Y}(\theta_{0,0})} & \gate{\mathrm{R_Z}(\theta_{0,1})} & \ctrl{1} & \gate{\mathrm{R_Y(\theta_{0,2})}} & \gate{\mathrm{R_Z(\theta_{0,3})}} & \qw \\
    	 	\nghost{\ket{{q}_{1}}} & \lstick{\ket{0}} & \gate{\mathrm{R_Y}(\theta_{1,0})} & \gate{\mathrm{R_Z}(\theta_{1,1})} & \targ & \gate{\mathrm{R_Y(\theta_{1,2})}} & \gate{\mathrm{R_Z(\theta_{1,3})}} & \qw  \\
    \\ }}
    \caption{Quantum Circuit for parameterized ansatz with 2 qubits ($n=4$).}
    \label{fig:quantum-ansatz-2-qubits}
\end{figure}

We did VQLS experiments with 2, 3, and 4 qubits, that represents time series of length $n$ of 4, 9, and 16. For each size of experiment, we used corresponding ansatz. Ansatz is parametrized quantum circuit, whose parameters can be adjusted to generate wide range of different quantum transformations and quantum states. Example of ansatz for 2 qubits is depicted on Figure ~\ref{fig:quantum-ansatz-2-qubits}.

\subsubsection{Cost Function}

The goal of the VQLS is to find the optimal parameter vector $\varphi$ for the ansatz state $\ket{\psi} = M \ket{x(\varphi)}$, where $\ket{x(\varphi)}$ is the state after the ansatz and $M$ decomposition of input matrix. We define the cost function similarly to the one in \cite{bravo-prieto_variational_2023}, but with a slight modification
\[
    C = \langle \psi | \psi \rangle - \langle \psi | b \rangle \langle b | \psi \rangle
\]
Since $\ket{\psi}$ is a unit vector, the first term simplifies to 1, yielding:
\[
    C = 1 - |\langle b | \psi \rangle|^2
\]
This expression can be rewritten as:
\[
    \langle b | \psi \rangle = \langle b | M \ket{x(\varphi)} \rangle = \langle b |  \sum_i \alpha_i M_i | x(\varphi) \rangle = \sum_i \alpha_i \langle b | M_i | x(\varphi) \rangle.
\]

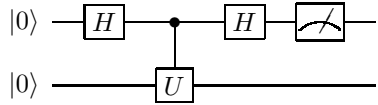
\begin{figure}[hbt!]
    \centering
    \scalebox{1}{
    \Qcircuit @C=1.2em @R=1.0em @!R {
        \\
        \nghost{\ket{{q}_{0}}} & \lstick{\ket{0} } & \gate{H} & \ctrl{1} & \gate{H} & \meter & \qw \\
        \nghost{\ket{{q}_{1}}} & \lstick{\ket{0}} & \qw & \gate{U} & \qw & \qw & \qw \\
        \\
    }}
    \caption{Hadamard Test circuit}
    \label{fig:hadamard-test}
\end{figure}

To evaluate each term in the sum, we use the Hadamard test. This means that the cost function is computed as a sum over all $M_i$, where for each term, the following procedure is performed: first, the ansatz state $\ket{x(\varphi)}$ is prepared; then, the corresponding $M_i$ transformation is applied. Afterward, the transformation that prepares the unit vector $b$ is applied to the state, and the resulting state is used as input for the Hadamard test as in Figure ~\ref{fig:hadamard-test}. The Hadamard test allows us to evaluate each term of $\langle b | \psi \rangle$. Finally, the result of the Hadamard test is multiplied by the corresponding $\alpha_i$, and the final cost function is obtained by summing over all terms.

As alternative to Hadamard test, Swap test can be used. Approach is similar to Hadamard test, and our experiments with Swap test have shown same outcomes as with Hadamard test, so we omit details of Swap test implementation.

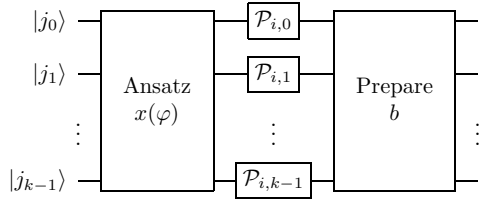
\begin{figure}[H]
    \centering
    \scalebox{0.85}{
    \Qcircuit @C=1.0em @R=0.8em @!R {
        \\
        \nghost{\ket{q_{0}}} & \lstick{\ket{j_0}} & \multigate{3}{\begin{array}{c}\text{Ansatz} \\ x(\varphi)\end{array}} & \gate{\mathcal{P}_{i,0}} & \multigate{3}{\begin{array}{c}\text{Prepare} \\ b\end{array}} & \qw \\
        \nghost{\ket{q_{1}}} & \lstick{\ket{j_1}} & \ghost{\begin{array}{c}\text{Ansatz} \\ x(\varphi)\end{array}} & \gate{\mathcal{P}_{i,1}} & \ghost{\begin{array}{c}\text{Prepare} \\ b\end{array}} & \qw \\
        \nghost{\vdots}      & \vdots             &  & \vdots &  & \vdots \\
        \nghost{\ket{q_{n}}} & \lstick{\ket{j_{k-1}}} & \ghost{\begin{array}{c}\text{Ansatz} \\ x(\varphi)\end{array}} & \gate{\mathcal{P}_{i,k-1}} & \ghost{\begin{array}{c}\text{Prepare} \\ b\end{array}} & \qw \\
        \\
    }}
    \caption{Quantum Circuit \( U \) for the Hadamard Test, shown in Figure~\ref{fig:hadamard-test}.}
    \label{fig:hadamard-test-U}
\end{figure}

The circuit in Figure ~\ref{fig:hadamard-test-U} illustrates the estimation of \( \langle b | M_i | x(\varphi) \rangle \) for a matrix \( M_i \), which represents the \( i \)-th term in the decomposition of \( M \) of Pauli gates \( \mathcal{P}_0, \dots, \mathcal{P}_{k-1} \). Initially, all $k = \log{n}$ qubits are prepared in the state \( \ket{0} \). A parameterized ansatz \( x(\varphi) \) is then applied, followed by the application of the \( M_i \) decomposition into Pauli gates $\mathcal{P}$. Finally, a transformation is applied to prepare the vector \( b \).

\section{Results}

\begin{figure}[H]
    \centering
    
    \begin{tikzpicture}
        \begin{axis}[
            width=\textwidth,
            height=0.5\textwidth,
            xlabel={Iteration},
            ylabel={Loss (MSE)},
            grid=major,
            axis lines=left,
            legend style={at={(0.98,0.98)}, anchor=north east},
            tick align=outside,
            xmin=0,
            ymin=0,
            scaled y ticks=false,
            y tick label style={/pgf/number format/fixed, /pgf/number format/precision=2},
        ]
            \addplot[
                color=blue,
                line width=0.8pt,
                unbounded coords=jump
            ]
            table[
                col sep=comma,
                header=true,
                x index=0,
                y index=1
            ] {data/fit_evals.csv};
            \addlegendentry{PQC with L-BFGS-B optimizer}

            \addplot[
                color=red,
                line width=0.8pt,
                dashed,
                unbounded coords=jump
            ]
            table[
                col sep=comma,
                header=true,
                x index=0,
                y index=2
            ] {data/fit_evals.csv};
            \addlegendentry{PQC with COBYLA optimizer}

            \addplot[
                color=green,
                line width=0.8pt,
                dash dot,
                unbounded coords=jump
            ]
            table[
                col sep=comma,
                header=true,
                x index=0,
                y index=3
            ] {data/fit_evals.csv};
            \addlegendentry{Classical linear model}

            \addplot[
                color=purple,
                line width=0.8pt,
                dotted,
                unbounded coords=jump
            ]
            table[
                col sep=comma,
                header=true,
                x index=0,
                y index=4
            ] {data/fit_evals.csv};
            \addlegendentry{Classical neural network}
        \end{axis}
    \end{tikzpicture}

    \caption{Loss function (MSE) comparison of different optimization methods and models over iterations. PQC with L-BFGS-B optimizer converged after 291 iterations.}
    \label{fig:mse-comparison}
\end{figure}
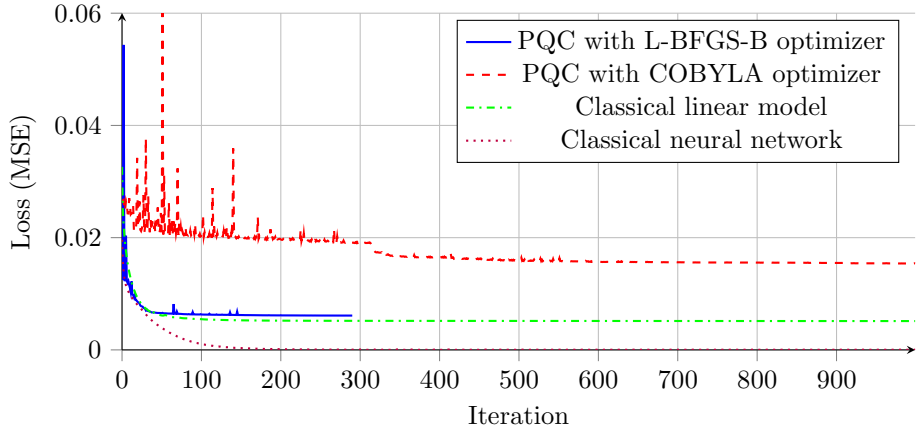

\begin{figure}[hbt!]
    \centering

    \begin{subfigure}{\textwidth}
        \centering
        \begin{tikzpicture}
            \begin{axis}[
                width=0.9\textwidth,
                height=0.3\textwidth,
                ylabel={M\euro},
                date coordinates in=x,
                xtick={2019-07-01,2020-01-01,2020-07-01,2021-01-01,2021-07-01,2022-01-01,2022-07-01,2023-01-01,2023-07-01,2024-01-01,2024-07-01},
                xticklabels=\empty,
                grid=major,
                xmin=2019-01-01,
                xmax=2024-08-01,
                scaled y ticks=false,
                legend to name=sharedlegend, 
                legend columns=5,
                legend style={draw=none, font=\small, legend cell align=left}
            ]
                \addplot[color=gray, solid, line width=0.5pt]
                table[col sep=comma, header=true, x index=0, y index=1] {data/model_predictions.csv};
                \addlegendentry{Data}
                \addplot[color=blue, solid, line width=1pt]
                table[col sep=comma, header=true, x index=0, y index=2] {data/model_predictions.csv};
                \addlegendentry{L-BFGS-B}

                \addlegendimage{color=red, dashed, line width=1pt}
                \addlegendentry{COBYLA}
                \addlegendimage{color=green, dash dot, line width=1pt}
                \addlegendentry{Linear Model}
                \addlegendimage{color=purple, dotted, line width=1pt}
                \addlegendentry{Neural Network}
                
            \end{axis}
        \end{tikzpicture}
    \end{subfigure}

    \begin{subfigure}{\textwidth}
        \centering
        \begin{tikzpicture}
            \begin{axis}[
                width=0.9\textwidth,
                height=0.3\textwidth,
                ylabel={M\euro},
                date coordinates in=x,
                xtick={2019-07-01,2020-01-01,2020-07-01,2021-01-01,2021-07-01,2022-01-01,2022-07-01,2023-01-01,2023-07-01,2024-01-01,2024-07-01},
                xticklabels=\empty,
                grid=major,
                xmin=2019-01-01,
                xmax=2024-08-01,
                scaled y ticks=false
            ]
                \addplot[color=gray, solid, line width=0.5pt]
                table[col sep=comma, header=true, x index=0, y index=1] {data/model_predictions.csv};
                \addplot[color=red, dashed, line width=1pt]
                table[col sep=comma, header=true, x index=0, y index=3] {data/model_predictions.csv};
            \end{axis}
        \end{tikzpicture}
    \end{subfigure}

    \begin{subfigure}{\textwidth}
        \centering
        \begin{tikzpicture}
            \begin{axis}[
                width=0.9\textwidth,
                height=0.3\textwidth,
                ylabel={M\euro},
                date coordinates in=x,
                xtick={2019-07-01,2020-01-01,2020-07-01,2021-01-01,2021-07-01,2022-01-01,2022-07-01,2023-01-01,2023-07-01,2024-01-01,2024-07-01},
                xticklabels=\empty,
                grid=major,
                xmin=2019-01-01,
                xmax=2024-08-01,
                scaled y ticks=false
            ]
                \addplot[color=gray, solid, line width=0.5pt]
                table[col sep=comma, header=true, x index=0, y index=1] {data/model_predictions.csv};
                \addplot[color=green, dash dot, line width=1pt]
                table[col sep=comma, header=true, x index=0, y index=4] {data/model_predictions.csv};
            \end{axis}
        \end{tikzpicture}
    \end{subfigure}

    \begin{subfigure}{\textwidth}
        \centering
        \begin{tikzpicture}
            \begin{axis}[
                width=0.9\textwidth,
                height=0.3\textwidth,
                ylabel={M\euro},
                date coordinates in=x,
                xticklabel style={rotate=45, anchor=east},
                xticklabel={\year-\month},
                xtick={2019-07-01,2020-01-01,2020-07-01,2021-01-01,2021-07-01,2022-01-01,2022-07-01,2023-01-01,2023-07-01,2024-01-01,2024-07-01},
                grid=major,
                xmin=2019-01-01,
                xmax=2024-08-01,
                scaled y ticks=false
            ]
                \addplot[color=gray, solid, line width=0.5pt]
                table[col sep=comma, header=true, x index=0, y index=1] {data/model_predictions.csv};
                \addplot[color=purple, dotted, line width=1pt]
                table[col sep=comma, header=true, x index=0, y index=5] {data/model_predictions.csv};
            \end{axis}
        \end{tikzpicture}
    \end{subfigure}

    \centering
    \pgfplotslegendfromname{sharedlegend}

    \caption{Comparison of different model predictions of the month-on-month sales growth against the actual data.}
    \label{fig:model-predictions}
\end{figure}
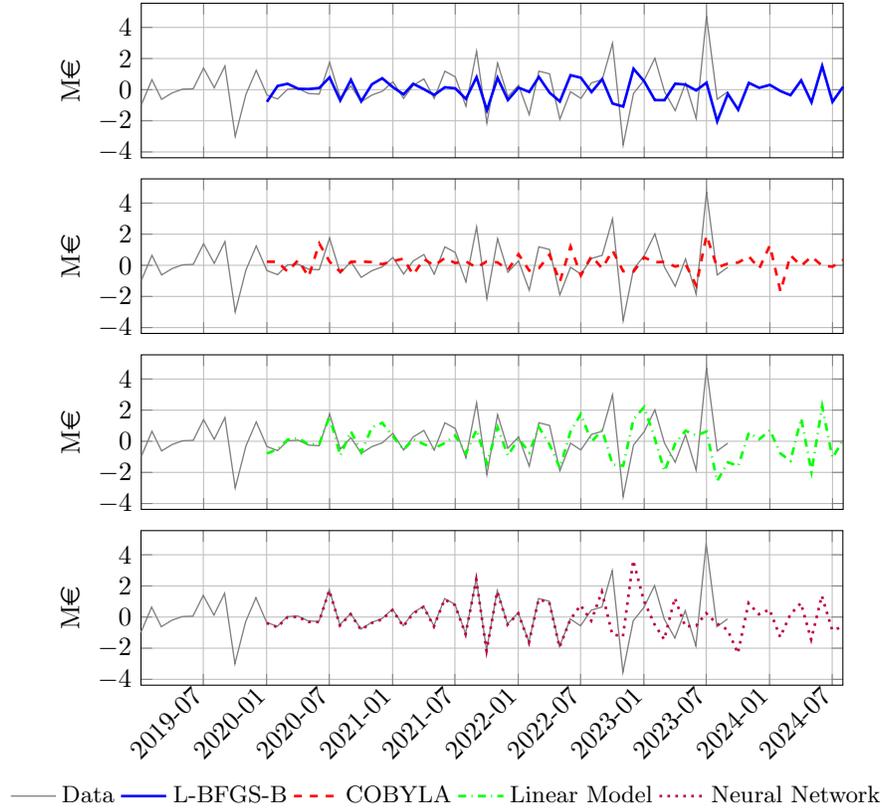

Figure~\ref{fig:mse-comparison} demonstrates the learning curve of different models, showing how training dataset MSE evolved over iterations. The PQC model using the L-BFGS-B optimizer showed the fastest convergence, reaching its minimum error at 291 iterations. In contrast, the COBYLA optimizer required more iterations and converged to a slightly higher error level. Among the classical models, the linear model achieved rapid convergence but displayed a higher final error compared to the quantum model with L-BFGS-B. The neural network showed fast improvement over iterations, leading to overfitting clearly visible in Figure~\ref{fig:model-predictions}. However, the PQC with L-BFGS-B outperformed all the classical models in terms of test set MSE, as summarized in Table~\ref{tab:mse_results}, indicating a potential advantage in using quantum-enhanced optimizers for this problem.

\begin{table}[hbt!]
\centering
\begin{tabular}{|l|c|c|}
\hline
\textbf{Model}           & \textbf{Training Set MSE} & \textbf{Test Set MSE} \\ \hline
PQC with COBYLA          & 0.01257   & 0.02106  \\ \hline
PQC with L-BFGS-B        & 0.00612   & 0.04418  \\ \hline
Classical Linear Model   & 0.00514   & 0.05177  \\ \hline
Classical Neural Network & 0.00003   & 0.05767  \\ \hline
\end{tabular}
\caption{Mean Squared Error (MSE) of models on training and test sets}
\label{tab:mse_results}
\end{table}
\pagebreak

The final MSE values on the test set, summarized in Table~\ref{tab:mse_results}, show that the PQC with COBYLA achieved the best performance on the test set, followed by the PQC with L-BFGS-B. However, both classical models were able to fit the training data better, and with proper regularization would likely lead to good test set MSE scores. No explicit regularization was applied to any of the models. The neural network, in particular, exhibited signs of overfitting, suggesting that techniques such as early stopping or other forms of regularization could significantly improve its performance.

Testing the VQLS algorithm with the provided matrix decomposition from \cite{OppenheimReznik} yielded unsatisfactory results. The time-series forecasting predictions were indistinguishable from random values, offering no reasonable forecast. It was also the case when we tested approach on real data from $Force$ $AI$. One key issue is the extremely high condition number of the corresponding arbitrary matrix. Transforming a custom matrix with $n$ rows and columns requires $O(n^2)$ Pauli matrices \cite{Werner}. An alternative solution for addressing this issue in future work could involve applying the Gram-Schmidt orthogonalization procedure. Another potential factor contributing to the poor performance is the COBYLA optimizer, which demands a large number of iterations to converge. On the other hand, results were satisfactory when we tried time series that strictly grow or strictly decline. In both cases, the output continued the growing/declining trend. This result may have potential uses in trading and investment areas in cases of growing and declining trends. Interestingly, we achieved ideal results when data represented geometrical progression (both increasing and decreasing), but it is quite artificial case to be considered useful practically. Overall, we think that this approach deserves further investigations, as it is based on shallow quantum circuits on small number of qubits, that is within capacity of available quantum computers.

Overall, these results suggest that the hybrid quantum-classical approach shows promise, but careful tuning of hyperparameters, such as the choice of optimizer, plays a crucial role and requires extensive experimentation. Once effective quantum models are identified, they could be integrated into ensembles of models, potentially aiding with regularization and complementing classical models by offering unique strengths and weaknesses, thereby enriching the ensemble's overall performance.

\section*{Acknowledgments}

The authors would like to express their gratitude to \emph{Forse AI} for providing the synthetic dataset used in this study and for their financial support, which made this research possible. 

\printbibliography

\end{document}